\documentclass[aps,prd,groupedaddress,twocolumn]{revtex4}

\usepackage{bbold}
\usepackage{url}
\usepackage{hyperref}
\usepackage[english]{babel}
\usepackage[utf8]{inputenc}
\usepackage{amsmath}
\usepackage{color}
\usepackage{amsfonts}
\usepackage{graphicx}
\usepackage{amssymb}
\usepackage{mathtools}
\usepackage{placeins}
\usepackage{float}
\usepackage{tabularx}
\usepackage{adjustbox}

\begin{document}

\title{Mode analysis of cosmological perturbations with the $E$-model $\alpha$-attractor inflaton potentials }

\author{Arunoday Sarkar}
\email{sarkararunoday7@gmail.com}
\author{Chitrak Sarkar}
\email{chitraksarkar@gmail.com}
\author{Buddhadeb Ghosh}
\email{bghosh@phys.buruniv.ac.in}

\affiliation{Centre of Advanced Studies, Department of Physics, The University of Burdwan, Burdwan 713 104, India}

\begin{abstract}
We have carried out detailed $k$-mode analysis of single-inflaton slow-roll inflationary perturbations including quantum fluctuations by setting up non-linear coupled differential equations of inflaton field ($\phi$), its perturbation ($\delta\phi$) and the metric perturbation  (the Bardeen potential, $\Phi_B$), and calculated the number of e-folds ($N$), scalar spectral index ($n_s$), tensor spectral index ($n_h$), scalar power spectrum ($\Delta_s$), tensor power spectrum ($\Delta_h$), tensor-to-scalar ratio ($r$) and the Hubble parameter ($H$) for different $k$ values at the horizon crossing. In these calculations we have employed the $E$-model $\alpha$-attractor potentials which are found to display slow-roll behaviour. The values of $n_s$ and $r$ obtained by us are consistent with those given by the well-known universal $\alpha$-attractor formulae. We got $n_s = 0.956908$ and $r= 0.005571 $ at $k = 10^6$ Planck unit for the value of the potential parameter $n = 1$. These can be compared with the Planck-2018 data \textit{viz.}, $n_s = 0.9649\pm 0.0042$ at $68\%$ CL, $r<0.064$ at $95\%$ CL and ACT-2020 data \textit{viz.}, $n_s = 0.9691\pm 0.0041$ at $68\%$ CL.

\end{abstract}

\maketitle

\section{Introduction}
\label{sec:intro}
The slow-roll inflationary scenario in the early universe was first introduced by Linde \cite{Linde:1981mu} and Albrecht and Steinhardt \cite{Albrecht:1982wi} in order  to rectify the shortcomings associated with the originally-proposed inflationary picture of Guth \cite{Guth:1980zm} and to provide better solutions for the cosmological problems such as the flatness, the horizon, the homogeneity and isotropy and the magnetic monopole problems.
\par The main ingredient in the improved inflationary scenario of \cite{Linde:1981mu} and \cite{Albrecht:1982wi} was a finite-temperature version of the Coleman-Weinberg (CW) potential \cite{Coleman:1973jx} at the one-loop order for the scalar field (here, the $\it{inflaton}$ field) which arises from the quantum correction of the classical potential and which can induce symmetry breaking in a gauge field theory. Linde, in \cite{Linde:1981mu}, made use of the CW mechanism in the Grand Unified Theory (GUT) scale and demonstrated its efficacy in solving the cosmological problems. \par The new (slow-roll) inflationary scenario \cite{Linde:1981mu,Albrecht:1982wi} was later employed by Dolgov and Linde \cite{Dolgov:1982th} to study the baryon asymmetry generation, and by Linde to investigate the scalar field fluctuations \cite{Linde:1982uu} and the temperature dependence of the Higgs boson coupling constant \cite{Linde:1982tg} during inflation.  \par Later on, quite a number of inadequacies of the CW potential in the description of the inflationary scenario of the early universe have been pointed out: (i) The potential gives a large density fluctuations \cite{Shafi:1983bd} which is about $10^5$ times too large \cite{Pi:1984pv} in comparison with the predicted value. This large density fluctuation arises from large gauge coupling constant in the CW potential \cite{Pi:1984pv}, (ii) The effective coupling constant $\lambda(\phi)$ due to the quantum fluctuations in the scalar field $\phi$ in the Coleman-Weinberg type GUT is large enough \cite{Linde:1982uu} to spoil the slow-roll condition by making the journey across the flat region of the potential too fast \cite{Steinhardt:1984jj}, (iii) The flatness in the CW potential during the spontaneous breaking of the SU(5) symmetry may be affected for another reason \cite{Steinhardt:1984jj,Breit:1983pp}. Instead of just two vacua corresponding to SU(5) and SU(3)$\times$ SU(2)$\times$U(1) symmetries, there may be intermediate metastable vacua due to other symmetries, such as SU(4)$\times$U(1).\par The CW potential, having the above difficulties, belongs to the category of small-field models, characterized by $\Delta\phi<M_{Pl}$ \cite{Baumann:2009ds}.\par With the shortcomings of the CW slow-roll potential in the perspective, various other inflationary models have been proposed \cite{Albrecht:1984qt,Albrecht:1985yf,Holman:1984yj}.\par By considering a model of one-loop-order CW potential, of a singlet complex scalar field, weakly coupled to Abelian gauge fields, Albrecht and Brandenberger \cite{Albrecht:1984qt} have shown that an inflation with sufficient length is realizable with a small value of the gauge coupling constant and a very large value of the symmetry breaking scale. A numerical study later on by Albrecht,  Brandenberger and Matzner \cite{Albrecht:1985yf} confirmed the results of \cite{Albrecht:1984qt}.\par In \cite{Holman:1984yj}, a substantially different cosmological inflationary scenario based on supersymmetry (SUSY) was proposed, in which inflation takes place after the GUT symmetry breaking and before the SUSY symmetry breaking. In this formalism, a satisfactory inflaton potential based on $\mathcal{N}=1$ supergravity can be obtained which yields the desired density fluctuation of $\mathcal{O}(10^{-4})$ despite a low reheat temperature.\par

In recent times, a new type of inflationary models has been proposed based on the concept of \textit{`cosmological attractors'} \cite{Remmen:2013eja,Remmen:2014mia,Linde:2016uec,Fumagalli:2016sof,Dubinin:2017irq}. Conventionally, an attractor is a solution,  a dynamical system shows in time starting from wide range of initial conditions. In the cosmological context, it has been shown \cite{Remmen:2013eja} that it is possible to obtain an (apparent) attractor behaviour in the inflationary scenario in an effective phase space of `non-canonical' variables. On a broader note, the concept of `attractor' and its universality were indicated by S.Ferrara and R.Kallosh in \cite{Ferrara:1996dd} and \cite{Ferrara:1996um}, in the context of supersymmetric blackholes. Subsequently, the term `cosmological attractors' was employed to denote a group or class of potentials to paraphrase their universal properties (detailed discussions in Sec.\ref{sec:review}) \cite{Kallosh:2013hoa,Kallosh:2013maa,Kallosh:2013daa,Kallosh:2013tua,Ellis:2013nxa,Buchmuller:2013zfa,Linde:2014nna,Carrasco:2015rva} and they were studied to describe the early universe physics namely, initial conditions for inflation \cite{Carrasco:2015rva} and asymptotic properties of inflaton field \cite{Kallosh:2016gqp} etc. Cosmological attractors added a new dimension in the field of slow-roll inflationary models, which in turn opened a doorway to comprehend the genesis of our universe. Understandably, these attractor potentials have removed many of the drawbacks of the early slow-roll potentials.\par 
The purpose of this paper is to present a detailed mode analysis of inflationary perturbations based on the $\alpha$-attractors vis-$\grave{a}$-vis the recent experimental data.

Our goal here is to study the microphysical aspects of quantum fluctuations during inflation as well as to discover the scale-independence of relevant cosmological parameters (such as $n_s$ and $r$) in the evolution of the expanding universe. \par
In Sec.\ref{sec:review}, we highlight the main aspects of the $\alpha$-attractor models, relevent for our study.
In Sec.\ref{sec:linear_scalar_tensor_perturb}, we derive systematically the coupled mode equations of the inflaton field, its perturbation and the Bardeen potential and write the mode-dependent power spectra and spectral indices.
In Sec.\ref{sec:results}, we describe the results and discussion regarding the present study. Our calculations have been mostly done through MATHEMATICA.
In Sec.\ref{sec:conclusions}, we write some concluding remarks.

\section{A brief review of \texorpdfstring{$\alpha$}{alpha}-attractors in inflationary scenario}
\label{sec:review}
The journey in search of an appropriate slow-roll potential started in two independent directions- one, Starobinsky model, based on  $R^2$ curvature correction \cite{Starobinsky:1980te,Whitt:1984pd,Kofman:1985aw,Maeda:1987xf} and two, based on chaotic initial conditions \cite{Linde:1983gd}. Later, their conformal \cite{Barrow:1988xh} and superconformal generalisations \cite{Goncharov:1983mw,Kallosh:2013lkr,Cecotti:1987sa,Ketov:2010qz,Ketov:2012jt, Ketov:2012se,Ellis:2013xoa,Kallosh:2010xz,Kallosh:2010ug,Kallosh:2011qk,Kallosh:2013pby} revealed that, despite of their completely unrelated routes of origin, their predictions almost coincided dramatically
\cite{Kallosh:2013hoa,Kallosh:2013maa,Kallosh:2013daa,Kallosh:2013tua,Ellis:2013nxa,Buchmuller:2013zfa,Linde:2014nna,Carrasco:2015rva}.\par  

Any inflationary potential and its efficacy to produce perturbations or fluctuations in the inflaton field are judged by the scalar spectral index ($n_s$) and the tensor to scalar ratio ($r$). These parameters when calculated for the potentials, mentioned above, were found to be (during inflation at leading order of $1/N$, $\mathcal{ O}(1/N^2)\ll1$, $N$ being the number of e-foldings): \\
1) For all Starobinsky-like models:
\begin{equation}
n_s = 1-\frac{2}{N}
\label{eq:ns_starobinsky}
\end{equation}
\begin{equation}
r = \frac{12}{N^2}
\label{eq:r_starobinsky}
\end{equation}
2) For all chaotic models (of different powers $n$ of inflaton field):
\begin{equation}
n_s = 1-\frac{2}{N}
\label{eq:ns_chaotic}
\end{equation}
\begin{equation}
r = \frac{8n}{N}
\label{eq:r_chaotic}
\end{equation} 
    Subsequent studies included many other single and multi inflaton potentials, e.g.  Higgs inflation\cite{Bezrukov:2007ep,Garc_a_Bellido_2009,Bezrukov:2009db,Giudice:2010ka} $\sim \lambda (\phi^2 - v^2)^2, v\gg 1$, hilltop inflation\cite{Boubekeur_2005}, Higgs model coupled non-minimally with gravity\cite{Barvinsky_2009,Linde:2011nh} $\sim \frac{\xi}{2}\phi ^2 R-\frac{\lambda}{4}(\phi ^2 - v^2)^2, \xi\ll0$ in the limit $1+\xi v^2 \ll 1$, natural inflation\cite{Freese:1990rb}, hybrid inflation\cite{Linde_1994} and non-minimally coupled $\phi^4$ model\cite{Okada:2010jf}, satisfied either (\ref{eq:ns_starobinsky}) and (\ref{eq:r_starobinsky}) or (\ref{eq:ns_chaotic}) and (\ref{eq:r_chaotic}), thereby showing there universal nature. All these classes of models/potentials were regarded as `conformal and superconformal attractors' or simply `{\it{cosmological attractors}}' to specify their following properties \cite{Kallosh:2013hoa}:\\
1) They all have covergent loci in $n_s - r$ parameter space towards an attracting point (called, `{\it{attractor}}') $n_s \approx 0.96$. The value of $n_s$ is potential independent in slow-roll region and $r$ depends on potentials.\\
2) All potentials have identical slow-roll pleteau at $V\approx 1-\frac{3}{2N}\approx 0.975$.\\
3) This plateau shows a universal asymptotic behaviour near the boundary of moduli space of the inflaton field. \par 

Microscopic origin of the attractor phenomena was revealed by two consecutive investigations- first, by S. Ferrara, R. Kallosh, A. Linde and M. Porrati \cite{Ferrara:2013rsa} for $\mathcal{N}=1$ minimal supergravity model of vector multiplet inflaton and second its chiral version by A. Linde, R. Kallosh and D. Roest \cite{Kallosh:2013yoa}. It was found that, all the potentials, previously regarded as attractors were found to be special cases of single potential\cite{Carrasco:2015rva}: 
\begin{equation}
V(\phi) = V_0 (1-e^{-\sqrt{\frac{2}{3\alpha}}\phi})^{n}
\label{eq:single_pot}
\end{equation}
Here the parameter $\alpha$, characterizing various attractor potentials, was found to be the inverse curvature of $SU(1,1)/U(1)$ internal embedding K\"ahler manifold,
\begin{equation}
 \alpha = -\frac{2}{3\mathcal{R}_K}
\label{eq:kahler_manifold}
\end{equation}
In \cite{Kallosh:2013yoa} general $\alpha$-dependent expressions of $n_s$ and $r$ were derived which at various limits and values of $\alpha$ reduced to the same equations as Eqs. (\ref{eq:ns_starobinsky}) to (\ref{eq:r_chaotic}), except at small $\alpha$ limit, it was found $r = \frac{12\alpha}{N^2}$. These universal attractor phenomena, described by the term `{\it{$\alpha$-attractors}}' were interpreted as manifestation of geometrical deformation of the K\"ahler manifold. Thus the $\alpha$-attractors made a bridge between the inflationary potentials and the internal geometry of space-time.
\par 
Two extreme limits of $\alpha$ were reported as two attractor points in the  $n_s - r$ space interpolating between the Starobinsky-like models (at small $\alpha$ limit) and the chaotic models (at large $\alpha$ limit) in Ref. \cite{Kallosh:2014rga,Kallosh:2014laa}.

\section{Linear scalar and tensor perturbations in the spatially flat gauge and their mode analysis}
\label{sec:linear_scalar_tensor_perturb}
In this section we follow the formalism given in \cite{Baumann:2009ds} and carry out our analysis in terms of physical time $t$. We use the natural system of unit, where $c=\hbar = 1$.\par

\textit{\underline{The metric}}: The metric with scalar perturbation in the spatially flat gauge ($\Psi$=E=0) reads as,
\begin{eqnarray}\nonumber
ds^2 &=& g_{\mu\nu}dx^\mu dx^\nu\nonumber\\ &=& -(1+2{\Phi})dt^2 +2a(t)\partial_i B dx^i dt\nonumber\\ &&+a^2 (t)\delta_{ij} dx^i dx^j
\label{eq:metric_scalar}
\end{eqnarray}
The perturbed metric tensor $g_{\mu\nu}$ can be written as,
\begin{equation}
g_{\mu\nu}={g^{(0)}}_{\mu\nu} +h_{\mu\nu},
\label{eq:perturbed_metric_tensor}
\end{equation}
where ${g^{(0)}}_{\mu\nu}$ is the unperturbed or the Friedmann-Lemaitre-Robertson-Walker (FLRW)  metric and $h_{\mu\nu}$ is the perturbation metric. For later calculations it will be useful to write $g_{\mu\nu}$, ${g^{(0)}}_{\mu\nu} +h_{\mu\nu}$ and $h_{\mu\nu}$ in matrix forms:

\begin{equation}
g_{\mu\nu}=
\begin{pmatrix}
-(1+2\Phi)&a\partial _1 B & a\partial_2 B & a\partial _3 B \\  
a\partial _1 B & a^2 & 0 & 0\\ 
a\partial_ 2 B & 0 & a^2 & 0 \\
a\partial _3 B & 0 & 0 & a^2 \\ 
\end{pmatrix},
\label{eq:gmunu_matrix}
\end{equation}
  
\begin{equation}
{g^{(0)}}_{\mu\nu}=
\begin{pmatrix}
-1 & 0 & 0 & 0 \\  
0 & a^2 & 0 & 0\\ 
0 & 0 & a^2 & 0 \\
0 & 0 & 0 & a^2 \\ 
\end{pmatrix},
\label{eq:g0munu_matrix}
\end{equation}
\begin{equation}
h_{\mu\nu}=
\begin{pmatrix}
-2\Phi &a\partial _1 B & a\partial_2 B & a\partial _3 B \\  
a\partial _1 B & 0 & 0 & 0\\ 
a\partial_ 2 B & 0 & 0 & 0 \\
a\partial _3 B & 0 & 0 & 0 \\ 
\end{pmatrix}
\label{eq:hmunu_matrix}
\end{equation}
The inverse metric can easily be obtained from (\ref{eq:gmunu_matrix}) as,
\begin{equation}
g^{\mu\nu}=
\begin{pmatrix}
-(1-2\Phi)& a^{-1}\partial ^1 B & a^{-1}\partial^2 B & a^{-1}\partial ^3 B \\  
a^{-1}\partial ^1 B & a^{-2} & 0 & 0\\ 
a^{-1}\partial^ 2 B & 0 & a^{-2} & 0 \\
a^{-1}\partial ^3 B & 0 & 0 & a^{-2} \\ 
\end{pmatrix}
\label{eq:gmunu_inverse}
\end{equation}
\\
\underline{\textit{The energy-momentum tensor}}:
The perturbed energy-momentum tensor is,
\begin{equation}
{T^\mu} _\nu (t,\vec{X})={{{T^{(0)}}^\mu}}_{\nu} (t) +\delta {T^ \mu} _\nu(t, \vec{X})
\label{eq:perturbed_energy_mom_tensor}
\end{equation}
where for a perfect fluid,
\begin{equation}
{T^\mu }_{\nu} = (p+\rho)u^\mu u_\nu + p {\delta ^\mu} _\nu.
\label{eq:perfect_fluid}
\end{equation} 
Consequently perturbations in ${T^\mu} _\nu$ is the result of perturbations in density, pressure and velocity :
\begin{equation}
\rho(t,\vec{X})= {\rho^{(0)}} (t) +\delta \rho (t, \vec{X}),
\label{eq:total density}
\end{equation}
\begin{equation}
p(t,\vec{X})= {p^{(0)}} (t) +\delta p (t, \vec{X}),
\label{eq:perturbation_dpt}
\end{equation}
and
\begin{equation}
u^ \mu (t,\vec{X})= {u^{(0)}}^\mu (t) +\delta u^\mu (t, \vec{X}).
\label{eq:velocity}
\end{equation}

In the Friedmann universe, the co-moving velocities of the perfect fluid can be written as,
\begin{equation}
{u^{(0)}}^\mu=(1,\vec{0}),  {u^{(0)}}_\mu=(-1,\vec{0})
\end{equation}
which satisfies the isotropy-constraint equation:
\begin{equation}
u^{\mu}u_{\mu}=-1.
\label{eq:isotropy_constraint}
\end{equation}

Eq.(\ref{eq:isotropy_constraint}) is assumed to be obeyed in the first-order perturbation also. Hence, using Eqs.(\ref{eq:velocity}) and (\ref{eq:isotropy_constraint}), we get,
\begin{equation}
\delta u_{0}=-\Phi.
\end{equation} 

We now define the physical velocity perturbation $v_{i}$ as,
\begin{equation}
\delta u_{i}= av_{i}.
\end{equation}
Then, writing,
\begin{equation}
u_{\mu}=(-1-\Phi, av_{i}),
\end{equation}
we can find using the metric (\ref{eq:gmunu_inverse}),
\begin{equation}
u^{\mu}=(1-\Phi, a^{-1}(v^{i}-\partial^{i}B)).
\end{equation}
We can now calculate the components of ${T^\mu }_{\nu}$ upto the first-order perturbation and set up the corresponding matrix structure as,

\begin{equation}
{T^\mu }_{\nu}=
\begin{pmatrix}
-\rho& \delta{\partial _1} q & \delta{\partial _2 }q &\delta{\partial _3} q  \\  
-a^{-2}\delta{\partial ^1} q+a^{-1}(p^{(0)}+\rho^{(0)}){\partial^1}B & p & 0 & 0\\ 
-a^{-2}\delta{\partial ^2} q+a^{-1}(p^{(0)}+\rho^{(0)}){\partial^2}B & 0 & p & 0\\ 
-a^{-2}\delta{\partial ^3} q+a^{-1}(p^{(0)}+\rho^{(0)}){\partial^3}B & 0 & 0 & p\\,  
\end{pmatrix}
\label{eq:energy_momentum_tensor}
\end{equation}
where, following the SVT decomposition\cite{Kurki-Suonio},  $\delta{\partial^i{q}}$  is the scalar perturbation part of a three-momentum density $\delta{q^i}$ defined as ,
\begin{equation}
\delta{q^i}=(p^{(0)}+\rho^{(0)})av^i.
\end{equation}
\\
\underline{\textit{The Christoffel symbols}}:
Upto the first-order perturbation following Christoffel symbols can easily be derived.
\begin{equation}
{{\Gamma}^0}_{ki}=(1-2\Phi)a^2H{\delta}_{ki}-a{\partial}_i {\partial}_kB,
\end{equation}
\begin{equation}
{{\Gamma}^0}_{00}=\dot{\Phi},
\end{equation}
\begin{equation}
{{\Gamma}^0}_{0i}={\partial}_i\Phi+aH{\partial}_iB,
\end{equation}
\begin{equation}
{{\Gamma}^i}_{00}=a^{-1}(H{\partial}^iB+{\partial}^i\dot{B})+a^{-2}{\partial}^i\Phi,
\end{equation}
\begin{equation}
{{\Gamma}^i}_{0j}=H{{\delta}^i}_j,
\end{equation}
and
\begin{equation}
{{\Gamma}^i}_{kj}=-aH{\partial}^iB{\delta}_{kj},
\end{equation}
where the Hubble parameter $H = \frac{\dot{a}}{a}$. It is function of time.\par 
\underline{\textit{The field equations}}:
We divide Einstein's equation,
\begin{equation}
{G^\mu}_\nu \equiv{ R^\mu}_\nu-\frac{1}{2}R{g^\mu}_\nu={T^\mu}_\nu
\end{equation}
into unperturbed and perturbation parts,
\begin{equation}
{{G^{(0)}}^\mu}_\nu+{\delta} {G^\mu}_\nu ={{T^{(0)}}^\mu}_\nu+{\delta} {T^\mu}_\nu.
\end{equation}
From the unperturbed Einstein's equation,
\begin{equation}
{{G^{(0)}}^\mu}_\nu={{T^{(0)}}^\mu}_\nu,
\end{equation}
and using (Eq.\ref{eq:g0munu_matrix}) in the metric ${{g^{(0)}}^\mu}_\nu$, we get the Friedmann equations,
\begin{equation}
H^2=\frac{\rho^{(0)}}{3}
\label{eq:Hubble_square}
\end{equation}
and
\begin{equation}
\dot{H}+H^2=\frac{\ddot{a}}{a}=-\frac{1}{2}(p^{(0)}+\frac{\rho^{(0)}}{3}).
\label{eq:Hubble_dot}
\end{equation}
Then combining Eqs.(\ref{eq:Hubble_square}) and (\ref{eq:Hubble_dot}), we get,
\begin{equation}
{\dot{\rho}}^{(0)}+3H(p^{(0)}+\rho^{(0)})=0.
\end{equation}
Now, the perturbation part in Einstein's equation gives, 
\begin{equation}
\delta {G^\mu}_\nu=\delta {T^\mu}_\nu.
\end{equation}
From the perturbations of the components of ${g^\mu}_\nu$ and ${T^\mu}_\nu$,  the following equations can be derived:
\begin{equation}
3H^2\Phi +\frac{{\vec{k}}^2}{a^2}(-aHB)=-\frac{1}{2}\delta\rho,
\label{eq:Bardeen}
\end{equation}
\begin{equation}
H\Phi=-\frac{1}{2}\delta q,
\label{eq:Bardeen2}
\end{equation}
\begin{equation}
H\dot{\Phi}+(3H^2+2\dot{H})\Phi=\frac{1}{2}\delta p,
\label{eq:Bardeen3}
\end{equation}
\begin{equation}
({\partial}_t+3H)\frac{B}{a}+\frac{\Phi}{a^2}=0.
\label{eq:Bardeen4}
\end{equation}
In Eq.(\ref{eq:Bardeen}), $\vec{k}$ is the three-momentum of the Fourier mode of the $B$-field, which will be the same as that of the perturbing quantized inflaton field, coupled to the Bardeen potential.\\

\underline{\textit{Coupled differential equations of density and}}\\ \underline{\textit{momentum perturbations}}:

Using the Eqs.(\ref{eq:Bardeen}), (\ref{eq:Bardeen2}), (\ref{eq:Bardeen3}), (\ref{eq:Bardeen4}), one can easily derive the following two coupled differential equations for the density and momentum perturbations.

\begin{equation}
\delta\dot\rho+3H(\delta{p}+\delta\rho)=\frac{{\vec{k}}^2}{a^2}\delta{q}-(p^{(0)}+{\rho}^{(0)}){\vec{k}}^2 \left(\frac{B}{a}\right)
\label{eq:coupled_diferential1}
\end{equation}
and
\begin{equation}
\delta\dot{q}+3H\delta{q}=-\delta{p}-(p^{(0)}+{\rho}^{(0)})\Phi.
\label{eq:coupled_diffrential2}
\end{equation}
One can also verify that Eqs.(\ref{eq:coupled_diferential1}) and (\ref{eq:coupled_diffrential2}) can be derived from the continuity equation,
\begin{equation}
{{T^{\mu}}_\nu};{\mu}=0.
\end{equation}
In order to obtain gauge-invariant physical results, we must now write Eqs.(\ref{eq:coupled_diferential1}) and (\ref{eq:coupled_diffrential2}) in terms of the Bardeen potentials\cite{Bardeen:1980kt}: $\Phi_B$ and $\Psi_B$. In the spatially flat gauge we can write,
\begin{equation}
\Phi_B=\Phi+\frac{d}{dt}(aB)
\label{eq:Bardeen_gauge1}
\end{equation}
and
\begin{equation}
\Psi_B=-aHB.
\label{Bardeen_gauge2}
\end{equation}
From Eqs.(\ref{eq:Bardeen4}), (\ref{eq:Bardeen_gauge1}) and (\ref{Bardeen_gauge2}), we get,
\begin{equation}
\Phi_B=\Psi_B.
\end{equation}
Using Eqs.(\ref{eq:Bardeen_gauge1}) and (\ref{Bardeen_gauge2}), we then write  Eqs.(\ref{eq:coupled_diferential1}) and (\ref{eq:coupled_diffrential2}) as,
\begin{equation}
\delta\dot\rho+3H(\delta{p}+\delta\rho)=\frac{{\vec{k}}^2}{a^2}\delta{q}+(p^{(0)}+{\rho}^{(0)})\frac{{\vec{k}}^2}{a^2} \left(\frac{\Phi_B}{H}\right)
\end{equation}
and
\begin{equation}
\delta\dot{q}+3H\delta{q}=-\delta{p}-(p^{(0)}+{\rho}^{(0)})\left[\Phi_B+\frac{d}{dt}\left(\frac{\Phi_B}{H}\right)\right].
\end{equation}
\\
\underline{\textit{Classical evolution equation of inflaton perturbation}} \\ \underline{\textit{minimally coupled to gravity}}: 
We shall now specialize the matter perturbation equations to the case of single-field inflaton perturbation. We shall be working in the Einstein frame of reference, thereby considering minimal coupling between the inflaton field and the gravitational field. The action is
\begin{equation}
S=\int d^{4}x\sqrt{-g}\left(\frac{1}{2}R+\mathcal{ L}_M\right)
\end{equation}
with the matter-field Lagrangian,
\begin{equation}
\mathcal{L}_M\equiv \mathcal{ L}_{\phi}=-\frac{1}{2}g^{\alpha\beta}{\partial}_{\alpha}\phi{\partial}_{\beta}\phi-V(\phi).
\end{equation}
Clearly, the action $S$ can be split up into the Einstein-Hilbert action $S_{EH}$ and the inflaton action $S_\phi$:
\begin{equation}
S=S_{EH}+S_\phi.
\end{equation}
Then from \cite{Kurki-Suonio}, the relation
\begin{equation}
T_{\mu\nu}(\phi)=-\frac{2}{\sqrt{-g}}\frac{\delta S_{\phi}}{\delta g^{\mu\nu}}
\end{equation}
yields,
\begin{equation}
T^{\mu}_{\nu}(\phi)=g^{\mu\lambda}\partial_\lambda\phi\partial_\nu\phi-{g^{\mu}}_\nu\left[\frac{1}{2}g^{\alpha\beta}\partial_\alpha\phi \partial_\beta\phi+V(\phi)\right].
\end{equation}
Using the perturbed metrics [Eqs.(\ref{eq:gmunu_matrix}) and (\ref{eq:gmunu_inverse})], we get,
\begin{equation}
{T^{0}}_{0}(\phi)=-\frac{1}{2}(1-2\Phi){\dot{\phi}}^{2}-\frac{1}{2}a^{-2}{\delta}^{ij}{\partial}_{i}\phi{\partial}_{j}\phi-V(\phi),
\label{eq:ground_energy}
\end{equation}
\begin{eqnarray}\nonumber
{T^{i}}_{j}(\phi) &=& \left(\frac{\partial^iB}{a}\right)\dot\phi\partial_j\phi+a^{-2}\partial^i\phi\partial_j\phi-{{\delta^i}_j} \left[-\frac{1}{2}(1-2\Phi){\dot\phi}^2\right.\nonumber\\&& +\left. \left(\frac{\partial^kB}{a}\right)\dot\phi\partial_k\phi+\frac{1}{2}a^{-2}\partial_k\phi\partial^k\phi+ V(\phi)\right],
\label{eq:pressure2}
\end{eqnarray}
\begin{equation}
{T^{0}}_{i}(\phi)=-(1-2\Phi)\dot\phi\partial_i\phi+\left(\frac{\partial^jB}{a}\right)\partial_j\phi\partial_i\phi,
\end{equation}
and
\begin{equation}
{T^{i}}_{0}(\phi)=\left(\frac{\partial^iB}{a}\right){\dot\phi}^2+a^{-2}(\partial^i\phi)\dot\phi.
\end{equation}
We we now introduce linear perturbation in the inflaton field:
\begin{equation}
\phi(t, \vec X)=\phi^{(0)}(t)+\delta\phi(t,\vec X)
\label{eq:linear_perturbation_inflaton}
\end{equation}
and obtain keeping upto the linear term, 
\begin{eqnarray}\nonumber
{T^{0}}_{0}(\phi)&=&\left[-\frac{{\dot\phi}^{(0)^2}}{2}-V(\phi^{(0)})\right]\nonumber\\&&+\left[-{\dot\phi}^{(0)}\delta\dot\phi+\Phi{\dot\phi}^{(0)^2}-\frac{\partial V}{\partial \phi^{(0)}}\delta\phi\right],
\label{eq:zeroth_order_energy}
\end{eqnarray}
\begin{eqnarray}\nonumber
{T^{i}}_{j}(\phi)&=&{\delta^i}_j \left[\frac{{\dot\phi}^{(0)^2}}{2}-V(\phi^{(0)})\right]\nonumber\\&&+{\delta^i}_j \left[{\dot\phi}^{(0)}\delta\dot\phi-\Phi{\dot\phi}^{(0)^2}-\frac{\partial V}{\partial \phi^{(0)}}\delta\phi \right],
\label{eq:pressure_perturbation2}
\end{eqnarray}
\begin{equation}
{T^{0}}_{i}(\phi)=-\partial_i({\dot\phi}^{(0)}\delta\phi),
\end{equation}
and
\begin{equation}
{T^{i}}_{0}(\phi)=\left(\frac{\partial^iB}{a}\right){\dot\phi}^{(0)^2}+a^{-2}\partial^i({\dot\phi}^{(0)}\delta\phi).
\end{equation}
Following Eqs. (\ref{eq:total density}), (\ref{eq:energy_momentum_tensor}),  (\ref{eq:linear_perturbation_inflaton}) and (\ref{eq:zeroth_order_energy}), we can identify the density perturbation due to the inflaton field as, 
\begin{equation}
\delta\rho(t,\vec X)={\dot\phi}^{(0)}\delta\dot\phi-\Phi{\dot\phi}^{(0)^2}+\frac{\partial V(\phi^{(0)})}{\partial \phi^{(0)}}\delta\phi.
\end{equation}
Similarly, from Eqs. (\ref{eq:perturbation_dpt}), (\ref{eq:energy_momentum_tensor}),  (\ref{eq:linear_perturbation_inflaton}) and (\ref{eq:pressure_perturbation2}), we get the pressure perturbation as,
\begin{equation}
\delta p(t,\vec X)={\dot\phi}^{(0)}\delta\dot\phi-\Phi{\dot\phi}^{(0)^2}-\frac{\partial V(\phi^{(0)})}{\partial \phi^{(0)}}\delta\phi.
\end{equation}
Also, in this way, we get the three-momentum density perturbation as,
\begin{equation}
\delta q=-{\dot\phi}^{(0)}\delta\phi.
\end{equation}
We can now convert the evolution equations of the density and three-momentum perturbations into the evolution equations of the inflaton perturbation and get,
\begin{equation}
\ddot\phi^{(0)}+3H\dot\phi^{(0)}+\frac{\partial V(\phi^{(0)})}{\partial \phi^{(0)}}=0
\label{eq:unperturbed_inflaton_equation}
\end{equation}
and
\begin{eqnarray}\nonumber
&& \delta\ddot\phi+3H\delta\dot\phi+\frac{{\partial}^{2} V(\phi^{(0)})}{\partial{ \phi^{(0)}}^{2}}\delta\phi+\frac{{\vec{k}}^2}{a^2}\delta\phi\\ && = \dot\phi^{(0)}
\left[\dot\Phi_B+\frac{d^2}{dt^2}\left(\frac{\Phi_B}{H}\right)+\frac{{\vec{k}}^2}{a^2}
\left(\frac{\Phi_B}{H}\right)\right]\nonumber\\&& -2\left[\Phi_B+\frac{d}{dt}\left(\frac{\Phi_B}{H}\right)\right]\frac{\partial V(\phi^{(0)})}{\partial \phi^{(0)}}.
\label{eq:perturbed_inflaton_equation}
\end{eqnarray}
Eqs.(\ref{eq:unperturbed_inflaton_equation}) and (\ref{eq:perturbed_inflaton_equation}) are respectively the evolution equations of the unperturbed inflaton field and the perturbation part of that field, both at the classical level. However, we should quantize the inflaton field with the idea that the spatially distributed quantum condensates are the seeds for the observed large scale structure of the universe which has developed through gravitational instabilities. Since we shall be working in the $\vec k$ space of the quantum inflaton field in an expanding background, we may write Eqs. (\ref{eq:unperturbed_inflaton_equation}) and (\ref{eq:perturbed_inflaton_equation}) in $\vec k$ space, considering the linear independence of the modes in this space:
\begin{equation}
\ddot\phi^{(0)}(k,t)+3H\dot\phi^{(0)}(k,t)+\frac{\partial V(\phi^{(0)}(k,t))}{\partial \phi^{(0)}(k,t)}=0,
\label{eq:momentum_inflaton}
\end{equation}
\begin{widetext}
\begin{eqnarray}
&&\delta\ddot\phi(k,t)+3H\delta\dot\phi(k,t)+\frac{{\partial}^{2}V(\phi^{(0)}(k,t))}{\partial{\phi^{(0)}(k,t)}^{2}}\delta\phi(k,t)+\frac{k^2}{a^2}\delta\phi(k,t)\nonumber\\&=&\dot\phi^{(0)}(k,t)\left[(\dot\Phi_B)(k,t)+\frac{d^2}{dt^2}\left(\frac{(\Phi_B)(k,t)}{H}\right)+\frac{k^2}{a^2}
\left(\frac{(\Phi_B)(k,t)}{H}\right)\right]\nonumber\\&&-2\left[(\Phi_B)(k,t)+\frac{d}{dt}\left(\frac{(\Phi_B)(k,t)}{H}\right)\right]\frac{\partial V(\phi^{(0)}(k,t))}{\partial \phi^{(0)}(k,t)}.
\label{eq:momentum_perturbation}
\end{eqnarray}
\end{widetext}
It should be noted that in Eqs.(\ref{eq:momentum_inflaton}) and (\ref{eq:momentum_perturbation}) we have dropped the three-vector notation on $k$ as these two equations depend only on the magnitude of $\vec{k}$,   as in the Mukhanov equation.\cite{Baumann:2009ds,Mukhanov:2005sc}\\

\underline{\textit{Quantization of the inflaton field:}} Eqs. (\ref{eq:momentum_inflaton}) and (\ref{eq:momentum_perturbation}) are valid for the mode functions $\phi(k,t)$ at the classical level.
They may belong to a quantum field when we quantize the inflaton field cannonically as,
\begin{equation}
\hat{\phi}(t,\vec{X})=\int \frac{d^3k}{(2\pi)^3}[\phi(k,t)\hat{a}(\vec{k})e^{i\vec{k}.\vec{x}}+{{\phi}^*}(k,t){\hat{a}^\dagger}(\vec{k})e^{-i\vec{k}.\vec{x}}],
\end{equation}
with the commutation of the annihilation and creation operators,
\begin{equation}
[\hat{a}(\vec{k}), {\hat{a}^\dagger}(\vec{k'})]=(2\pi)^3\delta(\vec{k}-\vec{k'}).
\end{equation}
The canonical momentum field can easily be calculated upto the first-order in perturbation to have the form,
\begin{equation}
\pi_\phi=\dot{\phi}^{(0)}+\delta\dot\phi-2\Phi\dot{\phi}^{(0)}.
\end{equation}
Here, $\phi(k,t)$ are the mode functions and the Eqs.(\ref{eq:momentum_inflaton}) and (\ref{eq:momentum_perturbation}), describing their time evolutions are valid both in the classical and the quantum formulations. An operator mode function can be written as,
\begin{equation}
\hat{\phi}(\vec{k},t)=\phi(k,t)\hat{a}_{\vec{k}}+\phi^{*}(-k,t){\hat{a}}^{\dagger}_{-{\vec{k}}}.
\label{eq:operator_inflaton}
\end{equation}

\underline{\textit{The power spectrum:}}
A two-point correlation function in a quasi de-Sitter universe is introduced  as the vacuum expectation value of a product of first-order perturbations in the quantum mode functions (Eq.(\ref{eq:operator_inflaton})) and can be expressed as,
\begin{equation}
\left\langle\delta\phi(\vec{k},t)\delta\phi(\vec{k'},t)\right\rangle =\frac{4\pi^3}{k^3}\delta(\vec{k}+\vec{k'})H^2
\end{equation}
where, on the right-hand-side the time dependence comes through the Hubble rate $H$. 
\par In the spatially flat gauge, we can relate the curvature perturbation $\mathcal{R}$ to the inflaton perturbation $\delta\phi$ as, 
\begin{equation}
\mathcal{R}=H\frac{\delta\phi}{\dot\phi}.
\end{equation} 
\par The quantum fluctuations in the inflaton field are microscopically induced in the curvature and one can write a curvature correlation function in terms of the inflaton field as,
\begin{equation}
\left\langle\mathcal{R}(\vec{k},t)\mathcal{R}(\vec{k'},t)\right\rangle=\frac{H^2}{\dot\phi^2}\left\langle\delta\phi(\vec{k},t)\delta\phi(\vec{k'},t)\right\rangle.
\end{equation} 
A dimensionless curvature power spectrum may be defined as,
\begin{equation}
\Delta_\mathcal{R}(t)=\frac{H^4}{(2\pi)^2{\dot{\phi}^2}},
\label{eq:curvature_power}
\end{equation} 
with its relation to the curvature correlation function as,
\begin{equation}
\left\langle\mathcal{R}(\vec{k},t)\mathcal{R}(\vec{k'},t)\right\rangle=\frac{16\pi^5}{k^3}\delta(\vec{k}+\vec{k'})\Delta_\mathcal{R}(t).
\end{equation} 
Similarly, for tensor perturbation we can write a dimensionless power spectrum for two polarizations as,
\begin{equation}
\Delta_h(t)=\frac{2}{{M_{Pl}}^2}\left(\frac{H}{\pi}\right)^2.
\label{eq:tensor_power}
\end{equation} 
\par Now, the time-dependence of the power spectra can be converted to $k$-dependence if we use the horizon-crossing condition, $k=aH$, whereby we include all the $k$ modes crossing the shrinking horizon during the inflationary period at different time.
\par With this understanding, we write the two power spectra (scalar and tensor) (Eq.(\ref{eq:curvature_power}) and(\ref{eq:tensor_power})) as,
\begin{equation}
\Delta_s(k)= \frac{1}{8\pi^2{M_{Pl}}^2}{\left(\frac{H^2}{\epsilon}\right)}_{k=aH}
\label{eq:scalar_power}
\end{equation}
and
\begin{equation}
\Delta_h(k)=\frac{2}{\pi^2{M_{Pl}}^2}{\left(H^2\right)}_{k=aH}.
\end{equation} 
where, in Eq.(\ref{eq:scalar_power}), we have identified the curvature perturbation with scalar perturbation and used the relation,
\begin{equation} 
{\dot\phi}^2=2{M^2_{Pl}} \epsilon H^2,
\end{equation} where $\epsilon$ is the slow-roll parameter and $M_{Pl}( = \frac{1}{\sqrt{8\pi G}})$ is the reduced Planck mass. \par Henceforth, we shall do our calculations with $M_{Pl}=1$.
\\ \\
\underline{\textit{Mode equations in k-space:}}
For studying the power spectra for different modes in the $k$-space, we should have solutions of the mode functions in the $k$-space. With this aim, we shall first change the $t$-derivatives in Eqs.(\ref{eq:momentum_inflaton}) and (\ref{eq:momentum_perturbation}) into $k$-derivatives.
\par It may be noted here that the horizon-crossing condition,
\begin{equation}
k=aH=a\frac{\dot{a}}{a}=\dot{a}
\label{eq:horizon_exit}
\end{equation}
will imply that each point in  $k$-space will correspond to a  different expansion rate. Likewise, each value of
\begin{equation}
\dot{k}=\ddot{a}=-\frac{1}{2}\left(p+\frac{\rho}{3}\right)a
\end{equation}
will correspond to a different accleration rate.
\par Now, applying the slow-roll condition, 
\begin{equation}
\dot{\phi}^2<<V(\phi),
\label{eq:slow_roll_condition1}
\end{equation}
we obtain from Eqs.(\ref{eq:energy_momentum_tensor}), (\ref{eq:ground_energy}), (\ref{eq:pressure2}) and (\ref{eq:linear_perturbation_inflaton}) of linear perturbation in the inflaton field, 
\begin{equation}
\rho\approx V(\phi) 
\label{eq:slow_roll_condition2}
\end{equation}
and
\begin{equation}
p \approx- V(\phi).
\label{eq:slow_roll_condition3}
\end{equation}
Thus from Eq.(\ref{eq:horizon_exit}), we get,
\begin{equation}
\dot{k}=\frac{1}{3}aV(\phi)
\label{eq:slow_condition_4}
\end{equation}
Also, retaining upto the first-order term in the Taylor series expansion of $V(\phi)$, we can write, 
\begin{equation}
V(\phi)=V(\phi^{(0)})\left( 1+\frac{\partial}{\partial\phi^{(0)}}\ln({V(\phi^{(0)}})\delta\phi\right) .
\label{eq:slow_roll_condition5}
\end{equation}
The scale factor can be written in terms of $V(\phi)$ as,
\begin{equation}
a=\frac{k}{H}=k\sqrt{\frac{3}{\rho}}=k\sqrt{3}{(V(\phi))}^{-\frac{1}{2}},
\label{eq:slow_roll_condition6}
\end{equation}
where, we have used,
\begin{equation}
H^2=\frac{\rho}{3}.
\label{eq:slow_roll_condition7}
\end{equation}
We, then, get from Eqs.(\ref{eq:slow_roll_condition1} - \ref{eq:slow_roll_condition7}),
\begin{equation}
\dot{k}=k\sqrt{\frac{V(\phi^{(0)})}{3}}\left( 1+\frac{\partial}{\partial\phi^{(0)}}\ln\sqrt{V(\phi^{(0)})}\delta\phi\right) 
\end{equation}
and
\begin{equation}
H=\sqrt{\frac{V(\phi^{(0)})}{3}}\left( 1+\frac{\partial}{\partial\phi^{(0)}}\ln\sqrt{V(\phi^{(0)})}\delta\phi\right).
\end{equation}
Then, converting the $t$-derivatives into $k$-derivatives, denotiong the latter by primes and retaining terms linear in perturbation we get respectively from Eqs. (\ref{eq:momentum_inflaton}) and (\ref{eq:momentum_perturbation}):
\begin{eqnarray}\nonumber
&&k^2{\phi}''^{(0)}+k^2G_1{{\phi}'^{(0)}}^2+4k{\phi}'^{(0)}\nonumber\\&& +k^2(G_1\delta\phi'+G_2{\phi}'^{(0)}\delta\phi){\phi}'^{(0)}+6G_1(1-2G_1\delta\phi)=0\nonumber\\
\label{eq:mode_equation1}
\end{eqnarray}
and
\begin{eqnarray}\nonumber
k^2\delta\phi''+(k^2 G_1{\phi}'^{(0)}+4k)\delta\phi'+[1+6(G_2+2G_1^2)]\delta\phi\nonumber\\ = \Phi_B''[k^3{\phi}'^{(0)}]
+\Phi_B'[2k^2{\phi}'^{(0)}-k^3G_1{{\phi}'^{(0)}}^2-12kG_1]
\nonumber\\ +\Phi_B[k{\phi}'^{(0)}-k^2{{\phi}'^{(0)}}^2G_1-k^3{\phi}'^{(0)}(G_2{{\phi}'^{(0)}}^2+G_1{\phi}''^{(0)})\nonumber\\-12G_1+12k{G_1}^2{\phi}'^{(0)}]\nonumber\\
\label{eq:mode_equation2}
\end{eqnarray}
where,
\begin{equation}
G_n=\frac{\partial^n}{\partial\phi^{(0)n)}}\ln\sqrt{V(\phi^{(0)})},   n=1,2,3
\end{equation}
 \par An evolution equation for the Bardeen potential can be derived from the perturbed Einstein's equations under the spatially flat gauge and with no anistropic stress,
\begin{equation}
3H^2\Phi-k^2H\frac{B}{a}=-4\pi G\delta \rho,
\end{equation}
\begin{equation}
H\Phi=-4\pi G\delta q,
\end{equation}
\begin{equation}
H\dot{\Phi}+(3H^2+2\dot{H})\Phi=4\pi G \delta p,
\end{equation}
\begin{equation}
(\partial_t +3H)\left(\frac{B}{a}\right)+\frac{\Phi}{a^2}=0.
\end{equation}
Trading $\Phi$ and $B$ for the Bardeen potential $\Phi_B$ and using the Horizon crossing condition, we get,
\begin{equation}
H\partial_t\left(\Phi_B +\frac{\delta \rho}{2H^2}\right)-3H^2\left(\Phi_B + \frac{\delta \rho}{2H^2}\right)=3\left(\delta \dot{q} +\frac{1}{2}\delta p\right).
\end{equation}
Converting the $t$-derivatives to $k$-derivatives as before, we obtain the differential equation for $\Phi_B$ as,
\begin{widetext}
\begin{eqnarray}\nonumber
k^4{{\phi}'^{(0)}}^2 {\Phi''}_B+(2k^4 {\phi}'^{(0)}{\phi}''^{(0)}-2k^3 {{\phi}'^{(0)}}^2-k^4G_1 {{\phi}'^{(0)}}^3-2k){\Phi'}_B
+(6+k^3(2-3kG_1{\phi'}^{(0)}){\phi''}^{(0)}{\phi'}^{(0)}\nonumber\\-k^2(k^2G_2
{{\phi'}^{(0)}}^2-3kG_1{\phi'}^{(0)}+4){{\phi'}^{(0)}}^2)\Phi_B\nonumber\\=6k(k{\phi''}^{(0)}+kG_1{{\phi'}^{(0)}}^2+{\phi'}^{(0)})\delta\phi+k^3{\phi'}^{(0)}\delta\phi''+k(k^2{\phi''}^{(0)}+2k{\phi'}^{(0)}+6G_1)\delta \phi'+6kG_2{\phi'}^{(0)}\delta\phi\nonumber\\
\label{eq:mode_equation3}
\end{eqnarray}
\end{widetext}
\par Eqs. (\ref{eq:mode_equation1}), (\ref{eq:mode_equation2}) and (\ref{eq:mode_equation3}) are coupled differential equations  for the $k$-space evolution of the unperturbed inflaton field, $\phi^{(0)}$ and its perturbation, $\delta\phi$ and the metric perturbation, $\Phi_B$ with minimal coupling with gravity. The metric perturbation, $\Phi_B$ and its derivatives appear on the r.h.s of Eq.(\ref{eq:mode_equation2}), thus making a connection between the quantum fluctuation and the geometrical perturbation.This aspect will be reflected in the $k$-dependence of the calculated power spectra and the spectral indices through the solutions of Eqs.(\ref{eq:mode_equation1}), (\ref{eq:mode_equation2}) and (\ref{eq:mode_equation3}).\\

\underline{\textit{Cosmological parameters:}}
The number of e-folds corresponding to the $\alpha$-attractor E-model potential ($V(\phi)$) can be obtained from the standard formula:
\begin{equation}
N (\phi) = \int_{\phi_\text{end}}^{\phi} \frac{V(\phi)}{\frac{dV(\phi)}{d\phi}} d\phi.
\end{equation}
The above integration can be evaluated with $V(\phi)$  in Eq.(\ref{eq:single_pot}) with $\alpha =1$ (small $\alpha$ limit) and then we get the result (see Appendix \ref{sec:appendixA}),
\begin{equation} 
N(\phi) \approx \frac{3}{2n} e^{\sqrt{\frac{2}{3}}\phi}.
\label{eq:number_of_efolds}
\end{equation} 
From this we can now write a mode-dependent number of e-folds as
\begin{equation}
N(k)\equiv N(\phi(k))=\frac{3}{2n}e^{\sqrt{2/3}\phi(k)}
\label{eq:mode_hubble}
\end{equation}
where $\phi(k) = \phi^{(0)}(k)+\delta\phi(k)$.
Using Eq. (\ref{eq:number_of_efolds}), the first and second potential slow-roll parameters can be expressed as functions of mode dependent number of e-folds as (see Appendix \ref{sec:appendixB}),
\begin{equation}
\epsilon_V (N(k)) \approx \frac{3}{4N(k)^2}.
\label{eq:first_slow_roll_parameter}
\end{equation} and 
\begin{equation}
\eta_V (N(k)) \approx \left(\frac{3}{2N(k)^2}-\frac{1}{N(k)}\right).
\label{eq:second_slow_roll_parameter}
\end{equation}

 The mode-dependent scalar and tensor spectral indices may be written respectively as,
\begin{equation}
n_s(k)=1+\frac{d\ln\Delta_s(k)}{d\ln k}
\end{equation}
and
\begin{equation}
n_h(k)=\frac{d\ln\Delta_h(k)}{d\ln k}.
\end{equation}
and the tensor-to-scalar ratio as,
\begin{equation} 
r(k)=\frac{\Delta_h(k)}{\Delta_s(k)}.
\label{eq:tensor_scalar_ratio1}
\end{equation}
Following \cite{Baumann:2009ds}, we can relate the mode-dependent spectral indices and the tensor-to-scalar ratio to the corresponding slow-roll parameters as,
\begin{equation}
n_s(k)=1+2\eta_V(k)-6\epsilon_V(k),
\label{eq:scalar_index}
\end{equation}
\begin{equation}
n_h(k)=-2\epsilon_V(k)
\label{eq:tensor_index}
\end{equation}
and
\begin{equation}
r(k)=16\epsilon_V(k),
\label{eq:tensor_scalar_ratio}
\end{equation}
where, $\epsilon_V(k)$ and $\eta_V(k)$ are the mode-dependent potential slow-roll parameters. Using Eqs. (\ref{eq:first_slow_roll_parameter}) and (\ref{eq:second_slow_roll_parameter}) in (\ref{eq:scalar_index}),  (\ref{eq:tensor_index}) and (\ref{eq:tensor_scalar_ratio}) we get (see Appendix \ref{sec:appendixC}),
\begin{equation}
n_s (N(k)) = 1-\frac{2}{N(k)}
\label{eq:final_scalar_index}
\end{equation}

\begin{equation}
n_h (N(k)) = -\frac{3}{2N(k)^2}
\end{equation}
\begin{equation}
r = \frac{12}{N(k)^2}
\label{eq:final_tensor_scalar_ratio}
\end{equation}
Eqs. (\ref{eq:final_scalar_index}) and (\ref{eq:final_tensor_scalar_ratio}) are the mode-dependent versions of the corresponding universal equations for the $\alpha$-attractors \cite{Carrasco:2015rva} in small $\alpha$ limit.

\section{Results and Discussion}
\label{sec:results}
\par
\begin{figure}[H]
	\centering
	\includegraphics[width=0.9\linewidth]{"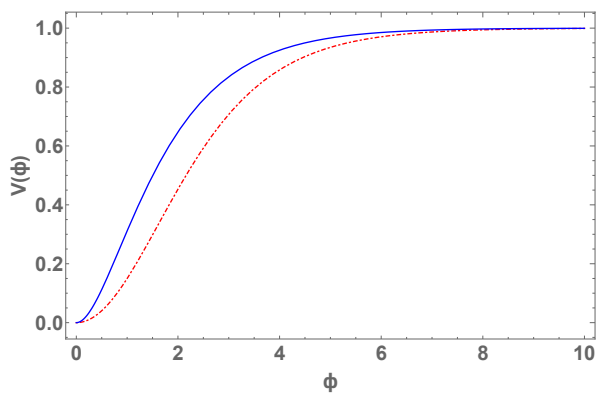"}
	\caption{Comparison of $\alpha $-attractor $T$-model potential ($V = V_0 \tanh^{n} \frac{\phi}{\sqrt{6\alpha}}$) (red, dot-dashed lower line) and the $E$-model potential ($V = V_0 (1-e^{-\sqrt{\frac{2}{3\alpha}}\phi})^{n} $) (blue, solid, upper line) for $\alpha =1, n = 2$, $V_0 = 1$ Planck unit and for positive $\phi$ values.}
	\label{fig:modeified-comparison} 
\end{figure}

 Although Eqs.(\ref{eq:mode_equation1}), (\ref{eq:mode_equation2}) and (\ref{eq:mode_equation3}) can, in general, be solved for various slow-roll inflaton potentials, in the present paper we consider the `universal $\alpha$-attractor' potentials \cite{Carrasco:2015rva} which are of latest interests and which have linkages with the higher scale physics such as supergravity and string theory\cite{Kallosh:2013yoa}. Also, out of the two main $\alpha$-attractor potentials viz., the $T$ model and $E$ model potentials, we have chosen the latter for our perturbative calculations as it has better slow-roll features as shown in fig.\ref{fig:modeified-comparison}. \par
\begin{figure}[H]
	\centering
	\includegraphics[width=0.9\linewidth]{"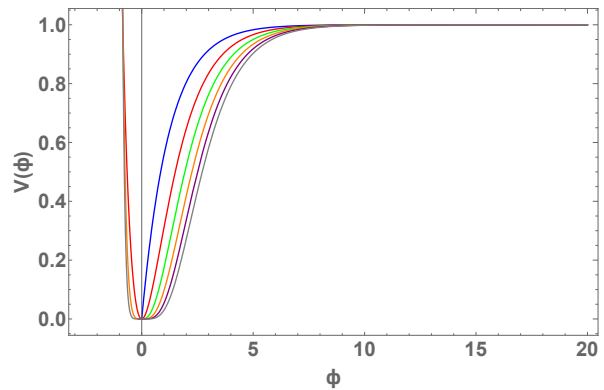"}
	\caption{$\alpha$-attractor E-model potentials, $V(\phi) = (1-e^{-\sqrt{\frac{2}{3\alpha}}\phi})^n$ vs. $\phi$,  for $\alpha = 1$ (maximum slow-roll feature) and $V_0 = 1$ Planck unit for various values of $n$. Here $n$ increases from left to right: $n=1$(blue), $n=2$(red), $n=3$(green), $n=4$(orange), $n=5$(purple) and $n=6$(gray).}
	\label{fig:potential}
\end{figure}

 In fig.\ref{fig:potential}, we show the slow-roll nature of the E-model $\alpha$ attractor potentials \cite{Carrasco:2015rva}, with which we have worked with in the present paper.
\par
\begin{figure}[H]
	\centering
	\includegraphics[width=0.9\linewidth]{"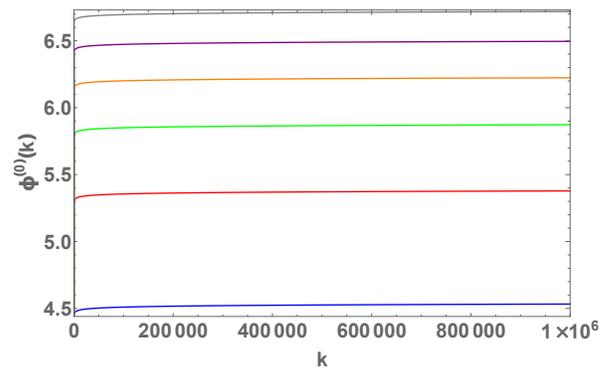"}
	\caption{$\phi^{(0)}(k)$ vs. $k$ obtained with slow-roll potentials shown in fig.\ref{fig:potential} with  $n=1$(blue), $n=2$(red), $n=3$(green), $n=4$(orange), $n=5$(purple) and $n=6$(gray) (from bottom to top.)}
	\label{fig:platinum-inflaton-jublee}
\end{figure}

\begin{figure}[H]
	\centering
	\includegraphics[width=0.9\linewidth]{"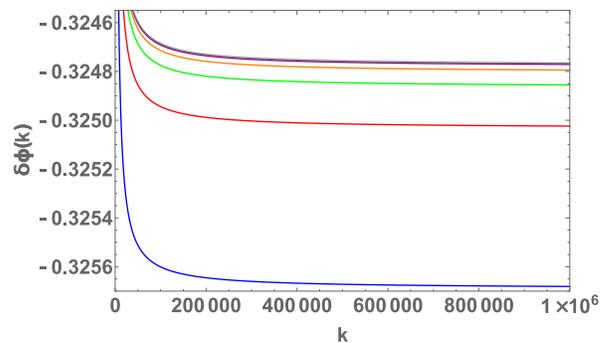"}
	\caption{$\delta\phi(k)$ vs. $k$ obtained with slow-roll potentials shown in fig.\ref{fig:potential} with  $n=1$(blue), $n=2$(red), $n=3$(green), $n=4$(orange), $n=5$(purple) and $n=6$(gray) (from bottom to top.) The results for $n=5$ and $6$ are same.}
	\label{fig:platinum-perturbed-jublee}
\end{figure}

\begin{figure}[H]
	\centering
	\includegraphics[width=0.9\linewidth]{"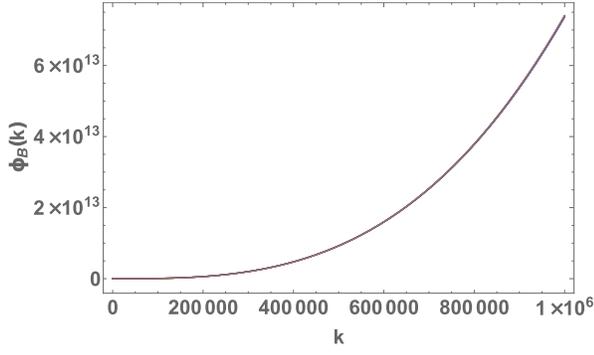"}
	\caption{$\Phi_{B}(k)$ vs. $k$ obtained with slow-roll potentials shown in fig.\ref{fig:potential} with  $n=1$(blue), $n=2$(red), $n=3$(green), $n=4$(orange), $n=5$(purple) and $n=6$(gray). Here results for different $n$ values are same.}
	\label{fig:platinum-bardeen-jublee}
\end{figure}

 In figs. \ref{fig:platinum-inflaton-jublee}, \ref{fig:platinum-perturbed-jublee} and \ref{fig:platinum-bardeen-jublee} we present our calculations of $\phi^{(0)}(k)$, $\delta \phi (k)$ and $\Phi_B (k)$. The initial values employed to obtain the solutions of the coupled nonlinear differential equations, here, are: $k = 1$ for $n = 1$ to $6$; $\phi^{(0)} = 4.518$ for $n=1$, $5.367$ for $n=2$, $5.863$ for $n=3$, $6.216$ for $n=4$, $6.489$ for $n=5$ and $6.712$ for $n=6$; $\phi'^{(0)} = 0$, $\delta\phi = 0.001$, $\delta\phi' = 0$, $\Phi_B = 0.001$, $\Phi'_B = 0$ for $n = 1$ to $6$.

 \par The large values of momenta ($ k= 4\times 10^5 \sim 10^6$ Planck unit) displayed in these diagrams connect to acausal regions outside the horizon .The results in fig.\ref{fig:platinum-perturbed-jublee} shows that the high momentum behaviour of the classical first order inflaton perturbation is qualitatively independent of the $n$ values of the slow-roll $\alpha$-attractor potentials. fig.\ref{fig:platinum-bardeen-jublee} displays large positive values of the Bardeen potential in the relevant range of the momentum, as it should be expected in the inflationary scenario. This result reflects the dominance of the large classical background against minimal variations of the inflaton potentials in the slow-roll regime. 

\begin{figure}[H]
	\centering
	\includegraphics[width=0.9\linewidth]{"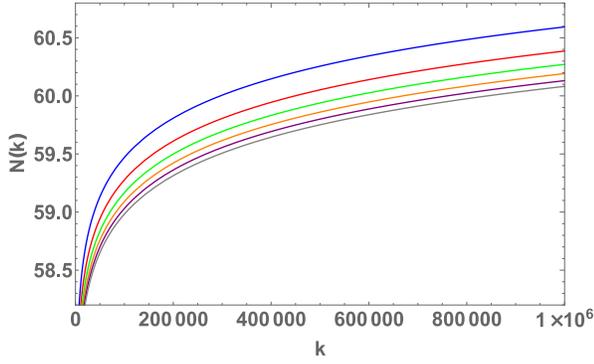"}
	\caption{Number of e-folds $N(k)$ vs. $k$ for the $\alpha$-attractor E-model potentials (fig.\ref{fig:potential}) for  $n=1$(blue), $n=2$(red), $n=3$(green), $n=4$(orange), $n=5$(purple) and $n=6$(gray) (from top to bottom).}
	\label{fig:number-of--e-folds}
\end{figure}

\begin{figure}[H]
	\centering
	\includegraphics[width=0.9\linewidth]{"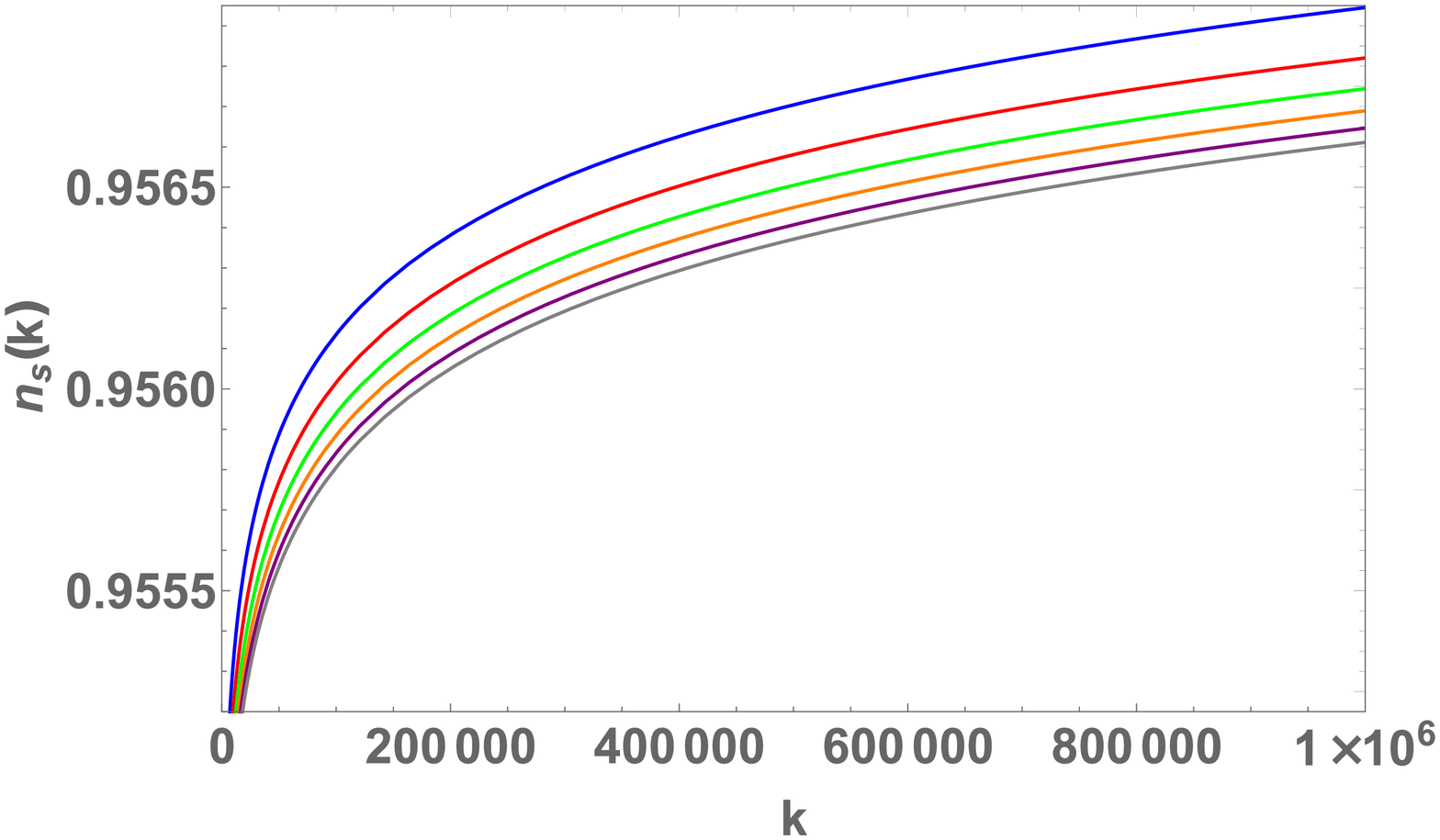"}
	\caption{Scalar spectral index ($n_s(k)$) vs. $k$ for the potentials in fig.\ref{fig:potential} for  $n=1$(blue), $n=2$(red), $n=3$(green), $n=4$(orange), $n=5$(purple) and $n=6$(gray) (from top to bottom.)  }
	\label{fig:scalar-spectral-index}
\end{figure}

\begin{figure}[H]
	\centering
	\includegraphics[width=0.9\linewidth]{"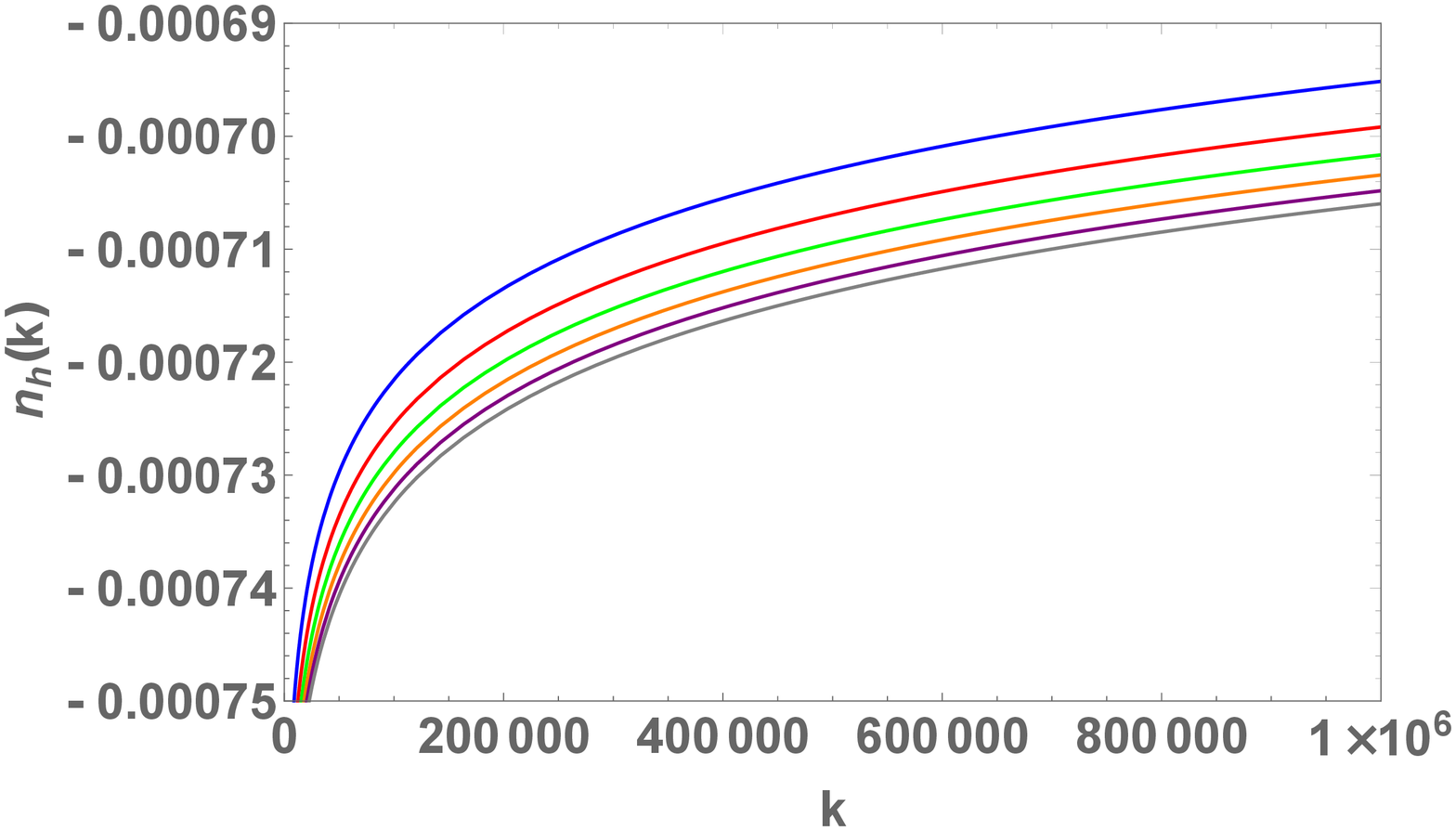"}
	\caption{Tensor spectral index $n_h (k)$ vs. $k$ for the potentials in fig.\ref{fig:potential} for $n=1$(blue), $n=2$(red), $n=3$(green), $n=4$(orange), $n=5$(purple) and $n=6$(gray) (from top to bottom.)}
	\label{fig:colour-tensor-spectral-index}
\end{figure}

\begin{figure}[H]
	\centering
	\includegraphics[width=0.9\linewidth]{"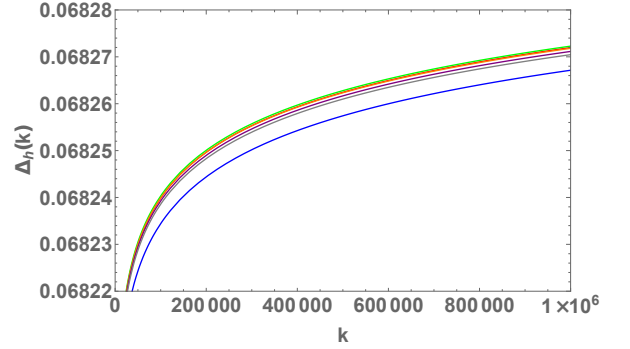"}
	\caption{Tensor power spectrum $\Delta_h (k)$ vs. $k$ for the potentials in fig.\ref{fig:potential} for  $n=1$(blue), $n=2$(red), $n=3$(green), $n=4$(orange), $n=5$(purple) and $n=6$(gray) (from bottom to top.) The results for $n=2$ through $6$ are same.}
	\label{fig:colour-tensor-power-spectrum}
\end{figure}
\begin{figure}[H]
	\centering
	\includegraphics[width=0.9\linewidth]{"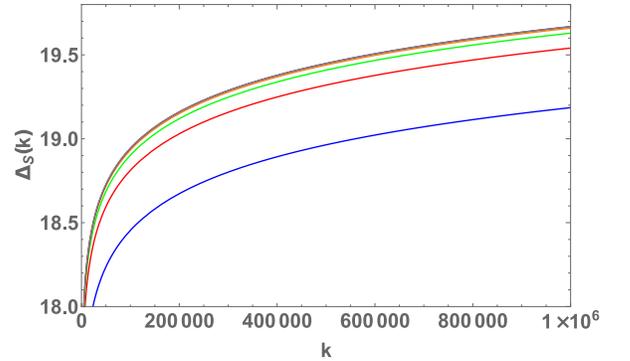"}
	\caption{Scalar power spectrum $\Delta_S (k)$ vs. $k$ for the potentials in fig.\ref{fig:potential} for  $n=1$(blue), $n=2$(red), $n=3$(green), $n=4$(orange), $n=5$(purple) and $n=6$(gray) (from bottom to top.) The results for $n=4$, $5$ and $6$ are same.}
	\label{fig:modified-scalar-power-spectrum}
\end{figure}

\begin{figure}[H]
	\centering
	\includegraphics[width=0.9\linewidth]{"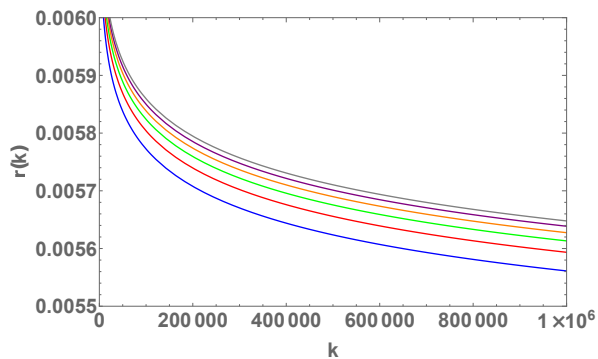"}
	\caption{Tensor to scalar ratio ($r(k)$) vs. $k$ for the potentials in fig.\ref{fig:potential} for  $n=1$(blue), $n=2$(red), $n=3$(green), $n=4$(orange), $n=5$(purple) and $n=6$(gray) (from bottom to top.) }
	\label{fig:tensor-to-scalar-ratio}
\end{figure}

\begin{figure}[H]
	\centering
	\includegraphics[width=0.9\linewidth]{"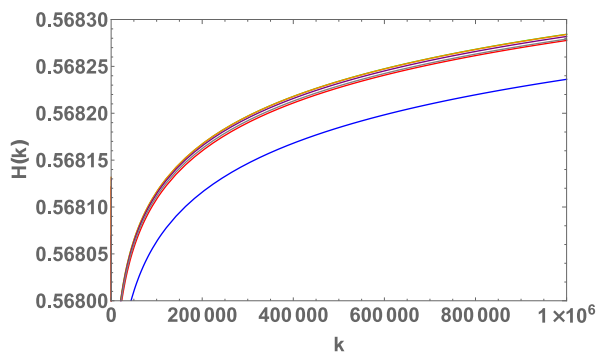"}
	\caption{Mode-dependent Hubble parameter $ H(k)$ vs. $k$ for the potentials in fig.\ref{fig:potential} for  $n=1$(blue), $n=2$(red), $n=3$(green), $n=4$(orange), $n=5$(purple) and $n=6$(gray) (from bottom to top.) The results for $n=2$ through $6$ are same.}
	\label{fig:colour-hubble-parameter}
\end{figure}

We have plotted the number of e-folds $N(k)$, the scalar spectral index $n_s (k)$, the tensor spectral index $n_h(k)$, the tensor power spectrum $\Delta_h(k)$, the scalar power spectrum $\Delta_s(k)$, the tensor-to-scalar ratio $r(k)$, and the Hubble parameter $H(k)$ vs. $k$ in figs. \ref{fig:number-of--e-folds}, \ref{fig:scalar-spectral-index}, \ref{fig:colour-tensor-spectral-index}, \ref{fig:colour-tensor-power-spectrum}, \ref{fig:modified-scalar-power-spectrum}, \ref{fig:tensor-to-scalar-ratio} and \ref{fig:colour-hubble-parameter} respectively for six values of the potential parameter $n$.\par Following Eq.(\ref{eq:mode_hubble}), we have plotted in fig.\ref{fig:number-of--e-folds} the number of mode-dependent e-folds, $N(k)$, vs. $k$ for various $n$ values of the $\alpha$-attractor potential. The plot shows saturations to values ($\sim 60$) which are in conformity with experiments\cite{Akrami:2018odb}. The figure also identifies the range of values of momenta, favourable for inflation.\par Table \ref{tab:table_name_parameters} shows the values of these parameters along with the number of e-folds for a representative large momentum viz., $k = 10^6$ Planck unit. The calculated values of $n_s$ and $r$ (particularly for $n=1$) can be compared with the corresponding experimental values, viz., $n_s = 0.9649 \pm 0.0042 $ at $68\%$ CL, $r < 0.064 $ at $95\%$ CL \cite{Akrami:2018odb} and $n_s = 0.9691\pm 0.0041$ at $68\%$ CL \cite{Aiola:2020azj}. So far as the Hubble rate is concerned we get $H=0.568236$ in Planck unit, which, after appropriate conversions of units \cite{Kolb:1990vq} comes out to be $1.6\times 10^{17}$ GeV. This is approximately $10^{60}$  times the experimental value, $H=66.88\pm0.92$ kms$^{-1}$ Mpc$^{-1}$\cite{Akrami:2018odb} $\cong 1.42\times 10^{-42}$ GeV, in the present universe. It may be noted that in a recent work \cite{Enqvist:2017kzh} the obtained upper limit of $H$ during inflation within the FIMP model of Dark Matter is $8.76\times 10^{13}$GeV.
\begin{table}[H]
\begin{center}
		\caption{\label{tab:table_name_parameters}}
		\begin{adjustbox}{width=0.5\textwidth}
		\begin{tabular}{ |c|c|c|c|c|c|c|c|c| } 	
			\hline
		    n &  $N$ & $n_s$ & $n_h$ & $\Delta_h$ & $\Delta_s$ & $r=\frac{\Delta_h}{\Delta_s}$ & \vtop{\hbox{\strut  $r$}\hbox{\strut (from}\hbox{\strut Eq.}\hbox{\strut \ref{eq:tensor_scalar_ratio1})}} & $H$ \\
			\hline 
			1 & 60.59 & 0.956908 & -0.000696 & 0.068267 & 19.1889 & 0.003558 & 0.005571 & 0.568236 \\ 
			2 & 60.39 & 0.956783 & -0.000700 & 0.068272 & 19.5439 & 0.003493 & 0.005603 & 0.568278 \\ 
			3 & 60.28 & 0.956707 & -0.000703 & 0.068272 & 19.6328 & 0.003477 & 0.005623 & 0.568284 \\ 
			4 & 60.20 & 0.956653 & -0.000705 & 0.068272 & 19.6627 & 0.003472 & 0.005637  & 0.568284 \\
			5 &  60.14 & 0.956610 & -0.000706 & 0.068271 & 19.6719 & 0.003470 & 0.005648 & 0.568282 \\
			6 & 60.09 & 0.956574 & -0.000707 & 0.068271 & 19.6722 & 0.003470 & 0.005657 & 0.568279 \\
			\hline
		\end{tabular}
		\end{adjustbox}
\end{center}
\end{table}

In the seventh and eighth columns of Table \ref{tab:table_name_parameters} we have compared the values of $r$ obtained from our calculations with those coming from the universal $\alpha$-attractor formula involving the number of e-foldings $N$. The fact that the difference in these two sets of values appears only in the third decimal place provides us with some confidence regarding our calculations.

\section{Conclusions}
\label{sec:conclusions}
In conclusion, we have studied inflationary perturbations with quantum fluctuations in the early universe within a single-inflaton model. Agreements of the obtained values of cosmological parameters viz., $n_s$, $r$ etc. with those of the experimental values reflect the fact that, the perturbations and fluctuations during inflation acted as seed in the evolution and formation of present universe whose signatures have remained in the background as cosmic gene. However there are many unresolved aspects regarding the upcoming investigations with which we may link up our present study:\\
1) A challenging task to the next frontier of the experiments is to explore the imprint of inflationary gravitational waves on CMB polarization viz., B-modes at $\geq 5\sigma$ level\cite{Abazajian:2013vfg}. Detection of those tensor modes is important to figure out the scale of inflation i.e. the super- and sub-Planckian nature of inflaton field, which will throw light on the UV cut off of quantum gravity and hence string theory.\\
2) The relic of those gravitational waves is parameterized through tensor-to-scalar ratio $r$, whose exact value is yet to be fixed. Planck puts an upper bound  $r < 0.064 $ at $95\%$ CL despite of its very feeble intensity \cite{Akrami:2018odb}. It will also constrain the inflationary Hubble parameter $H_I$.\par 
The methods of resolving the above puzzles will primarily hinge on relating the experimental observations to the primordial $k$-dependent power spectra \cite{Baumann:2009ds} and in this respect our work on mode analysis may provide some clue.
\section*{Appendix}
\label{sec:appendix}
\appendix
\section{Derivation of \texorpdfstring{$N(\phi)$}{Nphi} for \texorpdfstring{$\alpha$}{alpha}-attractor potentials}
\label{sec:appendixA}
The potentials for $\alpha$-attractor $E$-model with $\alpha=1$ is,
\begin{equation}
V(\phi)=V_0 (1-e^{-\sqrt{2/3}\phi})^n.
\end{equation}
\begin{equation}
\frac{dV(\phi)}{d\phi}=nV_0 \sqrt{2/3}(e^{-\sqrt{2/3}\phi})(1-e^{-\sqrt{2/3}\phi})^{n-1}.
\end{equation}
\begin{equation}
\frac{d^2 V(\phi)}{d\phi^2} = \frac{2nV_0}{3}(1-e^{-\sqrt{2/3}\phi})^n (e^{\sqrt{2/3}\phi}-1)^{-2}(n-e^{\sqrt{2/3}\phi}).
\end{equation}
The number of e-folds, then is
\begin{equation*}
N(\phi) = \int_{\phi_\text{end}}^{\phi}\frac{V(\phi)}{(dV(\phi)/d\phi)}d\phi
\end{equation*}
\begin{equation}
= \frac{3}{2n}[e^{\sqrt{2/3}\phi}-e^{\sqrt{2/3}\phi_\text{end}}-\sqrt{2/3}(\phi - \phi_\text{end})].
\label{eq:appendix4}
\end{equation}

First potential slow-roll parameter is
\begin{equation*}
\epsilon_V (\phi) = \frac{1}{2} \left(\frac{dV/d\phi}{V}\right)^2
\end{equation*}
\begin{equation}
= \frac{n^2}{3}(e^{\sqrt{2/3}\phi}-1)^{-2}.
\label{eq:appendix5}
\end{equation}
Inflation ends when $\epsilon_V (\phi_\text{end}) \approx 1$	. 
From this condition we get, 
\begin{equation}
\sqrt{2/3} \phi_\text{end} \approx \ln(1+\frac{n}{\sqrt{3}}).
\end{equation}
where we have considered only positive values of inflaton, i.e. $\phi_\text{end}>0$ (shown in table \ref{tab:table_name_inflaton_values}).\par 

\begin{table}[H]
		\begin{center}
		\caption{\label{tab:table_name_inflaton_values}}
		\begin{tabular}{ |c|c|c|c| } 	
			\hline
			n & $\sqrt{2/3}\phi_\text{end}$ & $e^{\sqrt{2/3}\phi_\text{end}}$ \\
			\hline 
			1	& 0.456 & 1.578 \\ 
			2	& 0.768 & 2.155 \\ 
			3   & 1.005 & 2.731\\
			4   & 1.197 & 3.102\\
			5   & 1.357 & 3.884\\
			6   & 1.496 & 4.464\\
			\hline
		\end{tabular}
	\end{center}
\end{table}
During inflation, inflaton rolls over infinitely long slow-roll plateau and appears in the downhill after the end of inflation. Therefore it is evident that (see Table \ref{tab:table_name_parameters}) the values of $\phi$ during slow roll are very large in comparison with $\phi_\text{end}$. Under this condition we can neglect the second and fourth terms in Eq.(\ref{eq:appendix4}) and take only the dominant exponential term. Then,
\begin{equation}
N(\phi)\approx \frac{3}{2n}e^{\sqrt{2/3}\phi}.
\label{eq:appendix7}
\end{equation}

\section{First and second potential slow-roll parameters}
\label{sec:appendixB}
Using Eqs. (\ref{eq:appendix5}), (\ref{eq:appendix7}) and considering the conditions, stated above, we get,
\begin{equation}
\epsilon_V (\phi) \approx \frac{n^2}{3}e^{-2\sqrt{2/3}\phi}
\label{eq:app_b_1}
\end{equation}
and therefore,
\begin{equation}
\epsilon_V(N)=\frac{3}{4N^2}.
\label{eq:app_b_2}
\end{equation}
The second potential slow-roll parameter is
\begin{equation}
\eta_V (\phi)= \left(\frac{d^2V/d\phi^2}{V}\right)=\frac{2n}{3}(e^{\sqrt{2/3}\phi}-1)^{-2}(n-e^{\sqrt{2/3}\phi}).
\end{equation}
Using the above mentioned equations and assumptions we obtain,
\begin{equation}
\eta_V (N)\approx\left(\frac{3}{2N^2}-\frac{1}{N}\right).
\label{eq:app_b_4}
\end{equation}

\section{Spectral indices and tensor-to-scalar ratio}
\label{sec:appendixC}
The scalar spectral index is
\begin{equation*}
n_s-1 = 2\eta_V - 6\epsilon_V.
\end{equation*}
Using (\ref{eq:app_b_2}) and (\ref{eq:app_b_4}) we can write,
\begin{equation*}
n_s (N)-1 = 2\eta_V(N) - 6\epsilon_V(N) =-\frac{2}{N}-\frac{3}{2N^2}.
\end{equation*}
Taking leading order of $1/N$ ($\mathcal{ O}(1/N^2)\ll 1$) we get,
\begin{equation}
n_s\approx1-\frac{2}{N}.
\end{equation}
From (\ref{eq:appendix7}) we can also write,
\begin{equation}
n_s (\phi) = 1-\frac{2}{N(\phi)} = 1-\frac{4n}{3}e^{-\sqrt{2/3}\phi}.
\end{equation}
The tensor spectral index is,
\begin{equation*}
n_h = -2\epsilon_V.
\end{equation*}
Using (\ref{eq:app_b_1}) and (\ref{eq:app_b_2}) we get,
\begin{equation}
n_h (\phi) = -2\epsilon_V(\phi)= -\frac{2 n^2}{3}e^{-2\sqrt{2/3}\phi}.
\end{equation}
and
\begin{equation}
n_h (N) = -2\epsilon_V(N) = -\frac{3}{2N^2}.
\end{equation}
The tensor to scalar ratio is,
\begin{equation*}
r = 16\epsilon_V.
\end{equation*}
Again from (\ref{eq:app_b_1}) and (\ref{eq:app_b_2}) we get,
\begin{equation}
r(\phi) =  16\epsilon_V(\phi) = \frac{16 n^2}{3}e^{-2\sqrt{2/3}\phi}
\end{equation}
and
\begin{equation}
r(N) =  16\epsilon_V(N) = \frac{12}{N^2}.
\end{equation}

\section*{Acknowledgement}
The present work has been carried out using some of the facilities provided by the University Grants Commission to the Center of Advanced Studies under the CAS-II program. A.S. and C.S. acknowledge the government of West Bengal for granting them the Swami Vivekananda fellowship and B.G. thanks Basundhara Ghosh for useful discussions.

\bibliography{biblio}

\end{document}